# Weak Anti-Localization Effect in Topological $Ni_3In_2S_2$ Single Crystal

Kapil Kumar[1,2], Yogesh Kumar[1,2] and V.P.S. Awana[1,2]

[1]*CSIR-National Physical Laboratory, Dr. K. S. Krishnan Marg, New Delhi-110012, India*
[2]*Academy of Scientific and Innovative Research (AcSIR), Ghaziabad 201002, India*

**Abstract:**

*$Ni_3In_2S_2$ is the most recent [1,2] entrant* into the family of topological insulator (TI) materials, the same exhibits very high MR in low temperature regime. Here, we report the crystal growth, structural, micro-structural, and magneto-transport study of $Ni_3In_2S_2$ down to 2.5K under applied field of up to 14Tesla. The phase purity and growth direction of single crystal is studied by performing XRD on both powder and flake, and further the Rietveld analysis is also carried out. The electrical transport measurements are studied and the grown crystal showed metallic behavior down to 2.5K, with $\rho_{300K}/\rho_{2.5K}$ ratio of around 7. A significant variation in magnetoresistance (MR) values is observed as the temperature is increased from 2.5K to 200K under applied field of up to 14Tesla. Interestingly the low T (2.5K), MR shows a clear v-type characteristic TI cusp. Magnetoconductivity data at low fields ($\pm$1Tesla) is fitted with Hikami Larkin Nagaoka (HLN) model, which showed the presence of weak anti-localization effect in the synthesized $Ni_3In_2S_2$ crystal at low temperatures. We have successfully grown near single phase $Ni_3In_2S_2$ and its TI behavior is demonstrated by magneto-transport measurements.

*Corresponding Author
Dr. V. P. S. Awana: E-mail: awana@nplindia.org
Ph. +91-11-45609357, Fax-+91-11-45609310
Homepage: awanavps.webs.com

## Introduction:

The class of ternary chalcogenides materials, with general formula $A_3B_2X_2$ crystallizes in the form of shandite structure, which contains three elements including a chalcogen [3]. Here A belongs to transition metals (Co, Ni), B is (In, Sn, Pb etc.) and X is (S & Se). In this article, we study the $Ni_3In_2S_2$, which belongs to sulfide family. The compound's chemical formula indicates the ratio of atoms 3:2:2 being present in its structure. The compound's properties are influenced by the combination of different elements present in its structure. Nickel, as a transition metal, contributes to its metallic and conductive characteristics. Indium, a post-transition metal, affects the electronic and structural properties of the compound. Sulfur, a non-metal, is responsible for the compound's chemical reactivity and bonding. These chalcogenides are also known to crystallize in Kagome lattice type structure [4]. $Ni_3In_2S_2$ is most recent member of $A_3B_2X_2$ family, which exhibits the Kagome lattice structure and has been the subject of extensive research in the field of condensed matter physics [1,2]. These materials are of interest because they can exhibit emergent topological states, such as Weyl cones [5,6], quantum spin liquids [7,8], quantum Hall states [9,10] and interesting magnetic properties [5,9,11-13]. When electrons reside on the corner-shared triangle network of the Kagome lattice, they create nontrivial quantum interference among the three sublattices. This interference leads to the formation of flat bands [14], saddle points [15], and Dirac fermions [16], which are characteristic excitations in Kagome metals. In addition to the characteristic excitations mentioned above, the presence of multi-bands near the Fermi level can give rise to topological nodal lines [2]. These nodal lines act as an avenue to realize topological semimetals [17,18], compensated semimetals [1] and Weyl semimetal [19]. The TIs are topic of interest as they possess unique properties such as high MR [20], Shubnikov de-Haas oscillations [21] and possibly the weak anti-localization effect (WAL). Previous reports show the presence of WAL effect in Kagome lattice type structures such as $Co_3SnS_2$ [19], but the WAL effect is not yet seen in $Ni_3In_2S_2$.

In this context we report the growth and characterization of $Ni_3In_2S_2$ single crystal to investigate insights into the rich physics of Kagome lattice type systems and their potential for hosting topological character through magnetoconductivity analysis. There have been only two reports [1,2] on the material under investigation $Ni_3In_2S_2$. Here, we show the PXRD Rietveld refinement characterization of the material and its resultant unit cell. In magneto-transport measurement, the observed MR is though less in comparison to previous reports [1,2], yet we found the WAL effect in $Ni_3In_2S_2$ through the HLN analysis. The fitted pre-factor parameter ($\alpha$) shows the dominance of topological surface states conduction at low temperature. The same TSS is predicted in the computational study of ref. 1 as well. These properties make Kagome-lattice compounds promising candidates for various electronic and spintronic applications, as well as for advancing our fundamental understanding of condensed matter physics.

**Experimental details:**

The $Ni_3In_2S_2$ single crystals were grown using a self-flux method [11,19]. The process involved weighing high-purity (99.999%) Nickel (Ni), Indium (In) and Sulphur (S) in the stoichiometric ratio, thoroughly mixing and grinding them in an inert atmosphere in an Argon-filled MBRAUN glove box. In order to prevent any oxidation during the heating process the resulting mixture was first pressed into rectangular pellets and then sealed into a quartz tube (with pressure $10^{-5}$ Torr). Figure 1, depicts the optimized heat treatment, which is followed in three steps to grow crystals by putting vacuum encapsulated tube in an auto-programmable furnace. Namely the three steps involved, heating up to 500°C, 700°C, and 950°C, with hold time of 48h, 48h and 24h, respectively followed by cooling to room temperature at a cooling rate of 60°C/h in first two steps. Whereas, in the third step the sample is slowly cooled down to 800°C with a cooling rate of 1°C/h and thereafter, the sample is air quenched. For further characterization, mechanically cleaved thin flakes of as synthesized crystals were extracted by using surgical blade. The X-ray diffraction (XRD) measurements were performed on both finely crushed powder and mechanically cleaved thin flakes by using Rigaku Miniflex-II X-ray diffractometer. The lattice parameters of $Ni_3In_2S_2$ single crystal were investigated by performing Rietveld refinement using Fullprof software. The surface morphology and elemental compositions were examined by performing scanning electron microscopy (SEM) and energy-dispersive X-ray analysis (EDAX) measurements using Zeiss EVO-50. Further to estimate the optical bandgap, we have performed UV-VIS spectroscopy on a UV 1800 Shimadzu spectrometer. The conventional four-probe arrangement is used to study the transport properties of synthesized crystals by using Quantum Design Physical Property Measurement System (QD-PPMS). For transport measurements, the sample is mounted on puck and electrical contacts were made using Epotek H20E silver epoxy. Magnetization studies were carried out using a Quantum Design MPMS (Magnetizatic Property Measurement System).

**Results and Discussion:**

Figure 2(a) displays the XRD pattern of mechanically cleaved thin flakes of obtained $Ni_3In_2S_2$ single crystal. The pattern shows high-intensity peaks corresponding to (*00l*) diffraction planes. This indicates that the crystal growth occurs along the c-axis, and the diffraction planes are stacked along (*003n*) where *n=1,2,3....* The inset of Figure 2(a) shows the image of obtained single crystal, which have a silvery shining surface. Further, to confirm the phase purity of the as-grown $Ni_3In_2S_2$ single crystal, XRD pattern was performed on finely crushed powder at room temperature in a 2θ range from 10° to 80°. To extract the lattice parameters of $Ni_3In_2S_2$, Rietveld refinement was performed as shown figure 2(b). The analysis revealed that the synthesized crystals belong to the $R\overline{3}m$ space group and have a rhombohedral crystal structure and peaks are indexed with their corresponding diffraction planes. In addition to single-phase of $Ni_3In_2S_2$ few unidentified diffraction peaks, denoted as "#," are also observed, which corresponds to the presence of Ni impurities. The lattice parameters obtained from the Rietveld refinement are $a = b$ = 5.375(1) Å and $c$ = 13.563(9) Å. The CIF (Crystallographic Information File) generated from the Rietveld refinement was used to draw the unit cell structure by using VESTA software as shown in figure 2(c). The unit cell structure illustrates that Ni-In layers form a Kagome structure and are sandwiched between S atoms. Also, there are total three such arrangements in a single unit cell. This is in agreement with the observed (*003n*) peaks seen in XRD pattern of mechanically cleaved crystal flake. Further, EDAX measurements were performed to determine the elemental composition of the $Ni_3In_2S_2$ single crystal. Figure 3(a) shows the observed EDAX spectra, and the inset table confirms the nearly stoichiometric elemental composition. The, SEM image (another inset) of the synthesized single crystal flake, reveals layered growth, confirming its crystalline nature.

The UV-VIS spectroscopy of the as-synthesized sample $Ni_3In_2S_2$ is displayed in Figure 3(b). The characterization approach known as UV-VIS makes use of the distinct absorption response to incoming UV light. As seen in Figure 3(b), The optical band gap of the synthesized $Ni_3In_2S_2$ can be calculated using the fundamental absorption spectra, which is equivalent to electron excitation from the valance band to the conduction band. Plotting $(\alpha h\nu)^2$ versus hν and extrapolating the straight part of the curve on the hν axis at α = 0 yields the direct band gap [22] of the sample, as indicated in Fig. 3(b). For $Ni_3In_2S_2$, obtained value of optical band gap is ~ 2.58eV.

Figure 4 shows the ρ-T plot in a temperature range from 300K to 2.5K, which clearly suggests the metallic behaviour for the synthesized $Ni_3In_2S_2$ single crystal. The RRR (residual resistivity ratio) is the ratio of the electrical resistivity at room temperature to the resistivity at lowest measured temperature. The obtained RRR value is nearly 7, which is lower than the previously reported [1] one, probably due to the presence of Ni impurity phase in our sample.

Figure 5 depicts the MR% versus applied transverse magnetic field of synthesized $Ni_3In_2S_2$ single crystal at temperatures of 2.5K, 50K, and 200K in a magnetic field range of ±14 Tesla. The formula is used to calculate the MR percentage (MR%) is given by

$$MR\% = \left[\frac{\rho(H) - \rho(0)}{\rho(0)}\right] \times 100 \qquad (1)$$

Here, $\rho$(H) is the resistivity under the influence of an applied magnetic field and $\rho$(0) is the resistivity with no applied magnetic field. It is observed that at 2.5K, the MR% is around 45%, indicating a high magnetoresistance, which is non-saturating up to highest measured magnetic field of 14 Tesla. Further, as the temperature is increased to 50K, the MR% decreases, and it becomes less pronounced (around 25%) and at 200K, the MR% further decreases to a small positive value of about 4%. In addition to non-saturated MR at higher fields a cusp like behaviour is also observed near zero magnetic field, which may be due to contribution from surface charge carriers in conduction mechanism of $Ni_3In_2S_2$ single crystal. However, as the temperature increases, the magnitude of MR% decreases, which can be due to bulk conducting channels being dominating the conduction mechanism.

The realisation of contribution of topological surface states in transport phenomenon were observed in MR measurements at low temperature, this evidence provokes to analysing the low field magnetoconductivity of $Ni_3In_2S_2$ for the expected weak anti-localization (WAL) effect. To confirm the presence of the WAL effect in the synthesized $Ni_3In_2S_2$ single crystal, the low-temperature magnetoconductivity (MC) is fitted using the HLN formula [18] which is given by

$$\Delta\sigma(H) = -\frac{\alpha e^2}{\pi h}\left[ln\left(\frac{B_\varphi}{H}\right) - \Psi\left(\frac{1}{2} + \frac{B_\varphi}{H}\right)\right] \qquad (2)$$

where, $\Delta\sigma(H) = \sigma(H) - \sigma(0)$, $\sigma(0)$ and $\sigma(H)$ are conductivity at zero and applied magnetic field, respectively. The characteristic field ($B_\varphi$) is given by the expression $B_\varphi = \frac{h}{8e\pi L_\varphi^2}$, here $h$ is Planck's constant, $e$ is the electronic charge, and $L_\varphi$ represents the phase coherence length. In the analysis of the HLN equation, the two fitting parameters are pre-factor $\alpha$ and phase coherence length $L_\varphi$. The pre-factor $\alpha$ determines the nature of localization present in the system which can be weak localization (WL) or WAL. A positive value of $\alpha$ signifies the weak localization (WL) effect, while a negative value of $\alpha$ indicates the presence of the weak antilocalization (WAL) effect. Also, the pre-factor $\alpha$ characterizes the presence of surface states in topological materials. For a single conducting channel, the value of $\alpha$ is expected to be -0.5 [17]. In materials with multiple topological surface states (TSS), the pre-factor $\alpha$ varies from -0.5. The value of $\alpha$ = -1 suggests the presence of two conduction channels in TIs [23-25]. Figure 6 (a)-(c), shows the magneto-conductivity vs. applied transverse magnetic field up to ±1Tesla at different temperatures. Here, the red symbols represent the experimental data points and the black solid lines represent the HLN fitted curve.

Table 2 represents the obtained values of pre-factor $\alpha$ and $L_\phi$ at different temperatures. At a temperature of 2.5K, as shown in Figure 6(a) the value of $\alpha$ is found to be -0.41, which is close to -0.5. This result suggests that a single TSS contributes to the conduction mechanism in the synthesized $Ni_3In_2S_2$ single crystal at this temperature. Further, as the temperature is increased to 50K and 200K (as shown in figures 6(b)&(c)) the value of $\alpha$ is found to be -0.07 and -0.04, respectively. The deviation of $\alpha$ from the standard values implies that the conduction in the synthesized $Ni_3In_2S_2$ single crystal is a result of the combined contribution of different conducting channels, which may include both TSS and bulk conducting states [26,27]. The $\alpha$ value extracted for the synthesized $Ni_3In_2S_2$ single crystal shows a similar trend as other topological materials possessing the WAL effect [28,29]. The observation of the WAL effect in $Ni_3In_2S_2$ matches with previously reported other Kagome lattice $Co_3Sn_2S_2$ [19]. To the best of our knowledge, this is the first report on the WAL effect of synthesized $Ni_3In_2S_2$ single crystal. Figure 6(d) shows the variation in the observed value of $\alpha$ and $L_\varphi$ as a function of temperature, which is extracted by performing HLN fitting.

To study the magnetic properties of $Ni_3In_2S_2$ single crystal, MPMS measurements were performed spanning a temperature range of 5-295K. Figure 7 depicts the M(T) curves obtained through the zero-field cooling (ZFC) and field cooling (FC) processes at a magnetic field strength of 250Oe. The two curves tend to merge near 295K indicating ferromagnetism, with a transition temperature nearby or above room temperature. The $M_{ZFC}(T)$ curve exhibits three distinct regions. At low temperatures (up to 12K) a slow increase, indicating the dominant paramagnetic nature at this region. Then up to 100K, a nearly linear increasing trend is observed, while above this temperature, the magnetization increases rapidly. On the other hand, the $M_{FC}(T)$ magnetization plot shows a pronounced decrease up to 12K (due to the paramagnetic phase) and then moderately decreases as the temperature is raised to 295K. The inset of figure 7 shows the M(H) graph at 5K, exhibiting the clear evidence of the dominance of ferromagnetism with the coercive field of 920(10)Oe. The observed ferromagnetism in the synthesized sample may be due to the presence of minor impurities of Ni. In agreement to the structural data, which is discussed earlier (PXRD-Figure 2b), the minor pure FM-Ni impurity is dominant, masking the inherent non-magnetic character of $Ni_3In_2S_2$ single crystal [2]. Being $Ni_3In_2S_2$ a new material, in future works the study can be extended to dielectric properties of the material following [30,31], and study in-plane magnetic anisotropy under different rotation speeds [32].

Overall, this study provides a comprehensive characterization of $Ni_3In_2S_2$ single crystals and demonstrates the presence of TSS and the WAL effect in this material. These findings contribute to the understanding of the electronic and magnetic properties of $Ni_3In_2S_2$ and may have implications for potential applications in spintronics and other electronic devices.

**Conclusion:**

In this study, single crystals of $Ni_3In_2S_2$ were synthesized using a 3-step solid-state reaction method, and the crystal is thoroughly characterized for its phase purity, morphology, and chemical composition. The electrical transport measurements revealed a positive MR at all measured temperatures, which gradually decreased as the temperature increased from 2.5K to 200K. The magnetization of studied $Ni_3In_2S_2$ is dominated by the minor impurity FM-Ni right up to room temperature. Interestingly, the analysis of HLN fitting clearly confirms the signatures of TSS and the presence of the WAL effect in synthesized $Ni_3In_2S_2$ single crystal.

**Acknowledgement:**

The authors are thankful to the Director of the National Physical Laboratory for his encouragement and keen interest in research activities. The authors would like to thank Prof. I. Felner from Hebrew Univ. Jerusalem for providing the magnetization measurements for our sample and acknowledge Dr. J. S. Tawale for FESEM measurements. Kapil Kumar is thankful to UGC, India for research fellowship and also thankful to AcSIR, India for Ph.D. registration. This research is funded by in house project no. OLP240832 and OLP240232.

**Table-1**

Unit cell parameters of synthesized crystal obtained from Rietveld refinement:

| Cell Parameters of **$Ni_3In_2S_2$** | Refinement Parameters |
|---|---|
| Cell type: Rhombohedral<br>Space Group: *R -3 m*<br>Lattice                          parameters:<br>$a$=$b$=5.3750(1)Å<br>& $c$=13.5639(6)Å<br>$\alpha$=$\beta$=90° & $\gamma$=120°<br>Cell volume: 339.3712Å$^3$<br>Density: 6.897gm/cm$^3$<br>Atomic co-ordinates:<br>Ni (0.5,0,0.5)<br>In1 (0,0,0)<br>In2 (0,0,0.5)<br>S (0,0,0.30761) | $\chi^2$=3.94<br>$R_p$=8.9916<br>$R_{wp}$=12.3986<br>$R_{exp}$=6.2392 |

**Table: 2**

Low field (up to ±1 Tesla) HLN fitted parameters of $Ni_3In_2S_2$ single crystal.

| Temperature (K) | $\alpha$ | $L_\varphi$ (nm) |
|---|---|---|
| 2.5 | -0.41 | 44.9(5) |
| 50 | -0.07 | 35.9(4) |
| 200 | -0.04 | 24.7(6) |

**Figure captions:**

**Figure 1:** Schematic of heat treatment followed to synthesize $Ni_3In_2S_2$ single crystal.

**Figure 2: (a)** XRD pattern taken on crystal flake of synthesized $Ni_3In_2S_2$ single crystal. **(b)** Rietveld refined XRD pattern of $Ni_3In_2S_2$ performed on crushed powder. **(c)** Unit cell structure of synthesized $Ni_3In_2S_2$ single crystal processed in VESTA software.

**Figure 3: (a)** EDAX spectra of synthesized $Ni_3In_2S_2$ single crystal, in which inset is showing the elemental composition of constituent elements and FESEM image of surface morphology of synthesized $Ni_3In_2S_2$ single crystal. **(b)** Calculation of optical band gap from UV–VIS spectroscopy of synthesized $Ni_3In_2S_2$.

**Figure 4:** Resistance versus temperature curve of synthesized $Ni_3In_2S_2$ single crystal.

**Figure 5:** MR% versus H(T) plots at 2.5,50 and 200K curve of synthesized $Ni_3In_2S_2$ single crystal.

**Figure 6:** HLN fitted low field (± 1 T) MC of synthesized $Ni_3In_2S_2$ single crystal at **(a)** 2.5 K **(b)** 50 K **(c)** 200 K **(d)** The variation of HLN fitting parameters with respect to temperature.

**Figure 7:** M-T plot of $Ni_3In_2S_2$ single crystal at 250Oe in both FC and ZFC, the inset shows the M-H plot obtained at 5K.

**Compliance with Ethical Standards Statement:**
All funding sources are duly acknowledged.
**Statement of Competing Interests:**
The authors have no competing interest.
**Research Data Policy and Data Availability Statements:**
Research data will be available from the corresponding author on reasonable request.
**Author Contribution:**
All the authors contributed equally to the MS.

Fig.1:

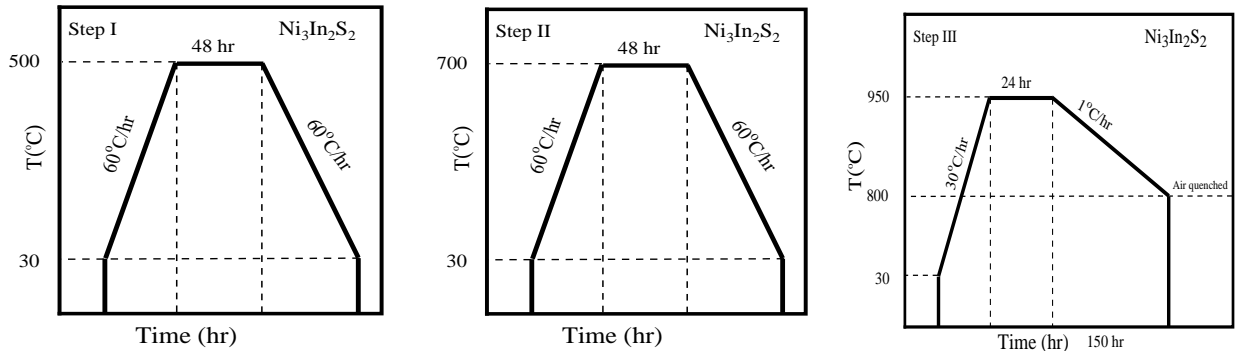

Fig. 2(a):

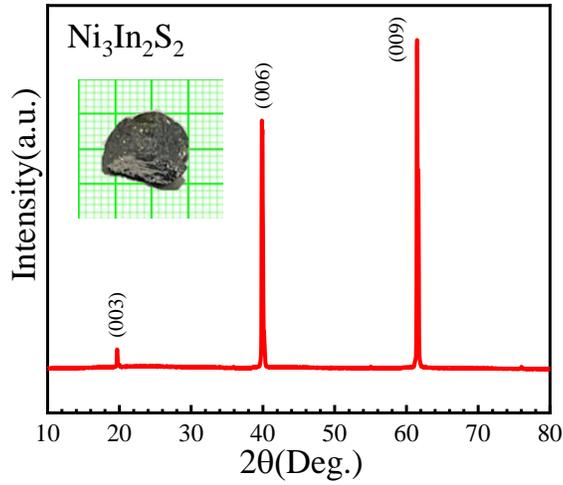

Fig. 2(b):

Fig. 2(c):

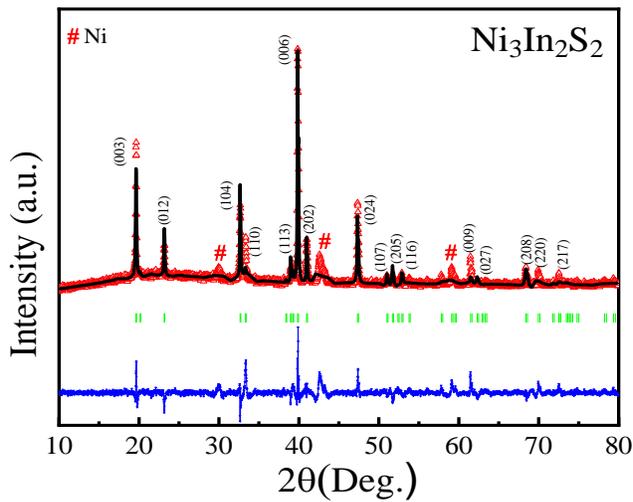

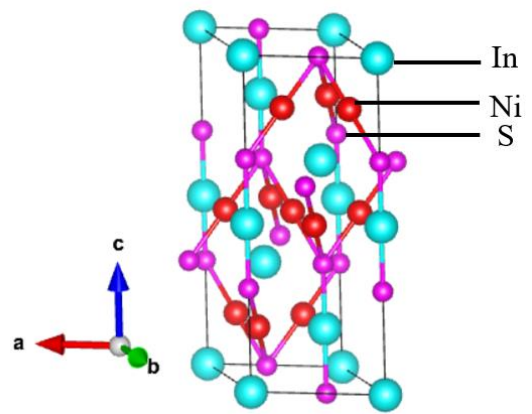

Fig. 3(a)

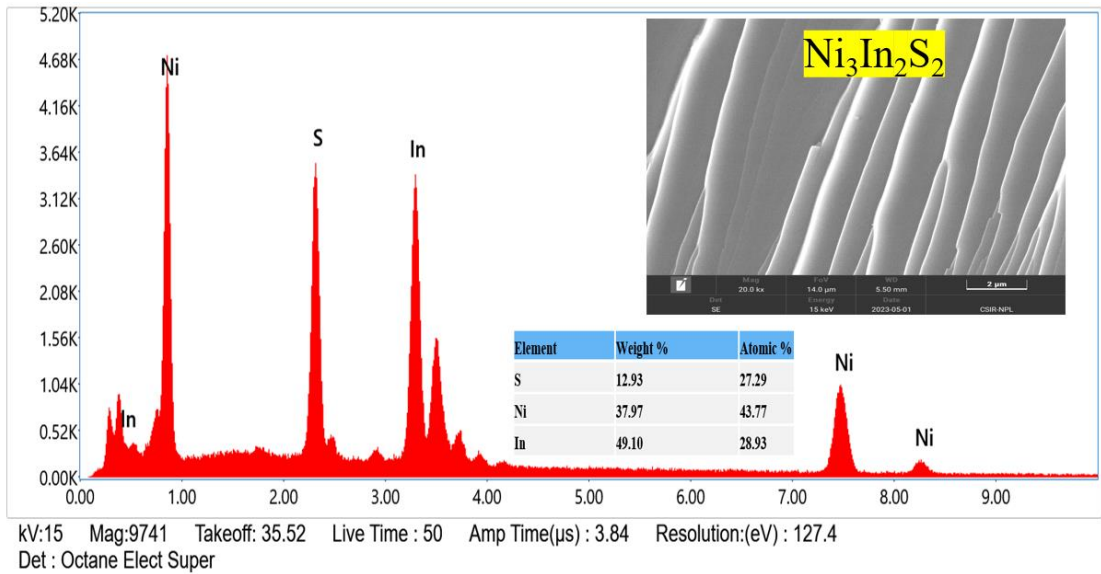

Fig. 3(b)

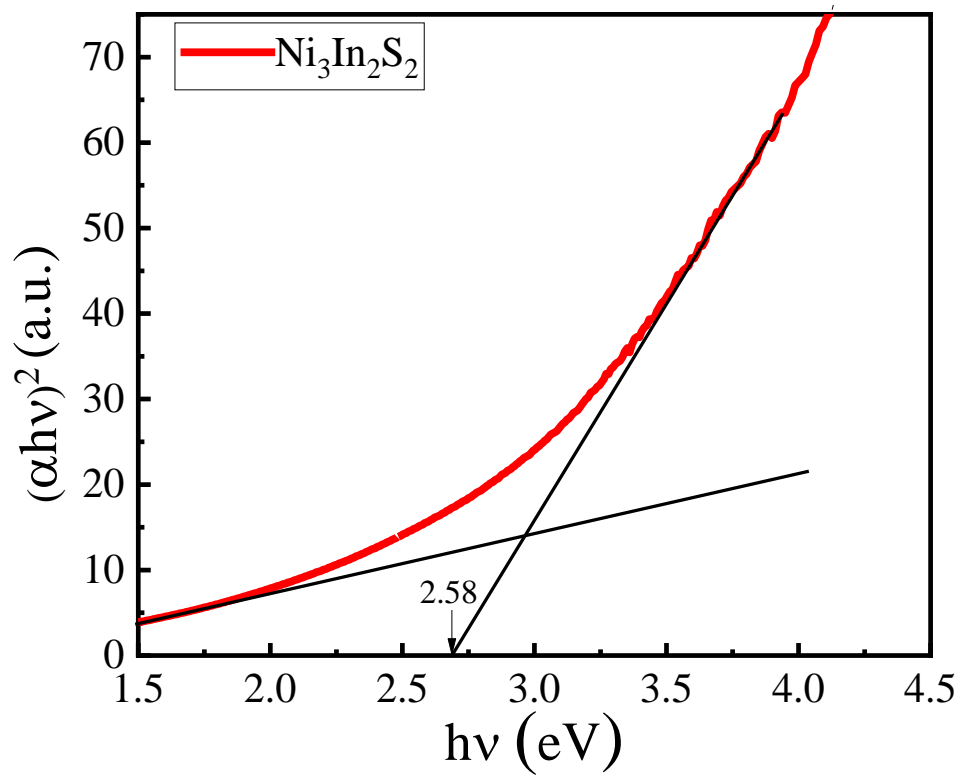

Fig. 4

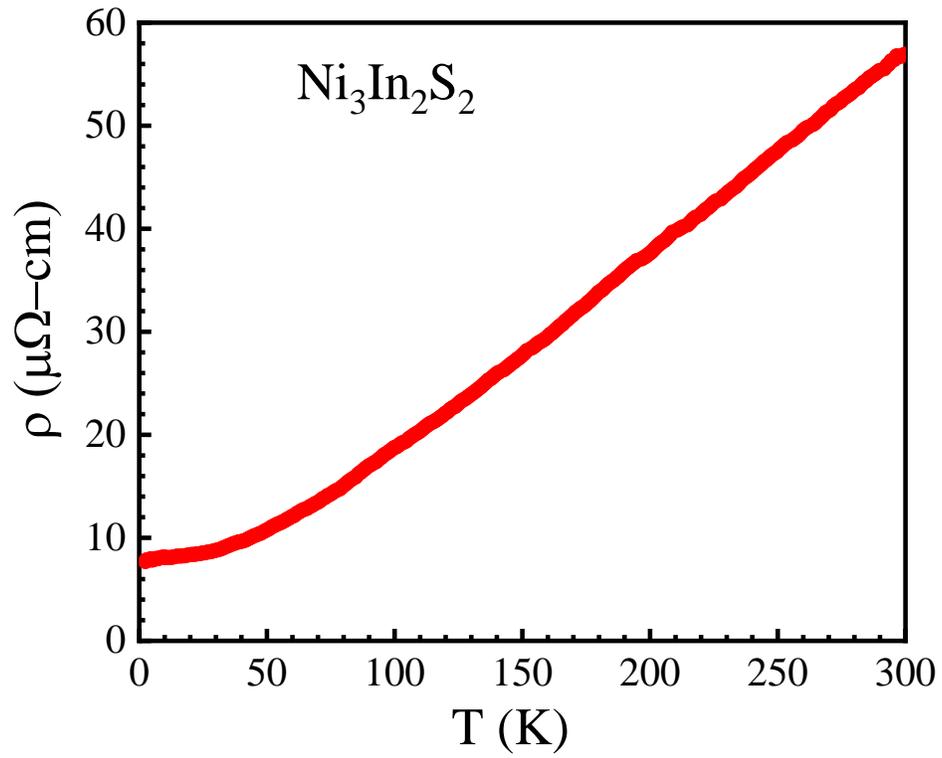

Fig.5:

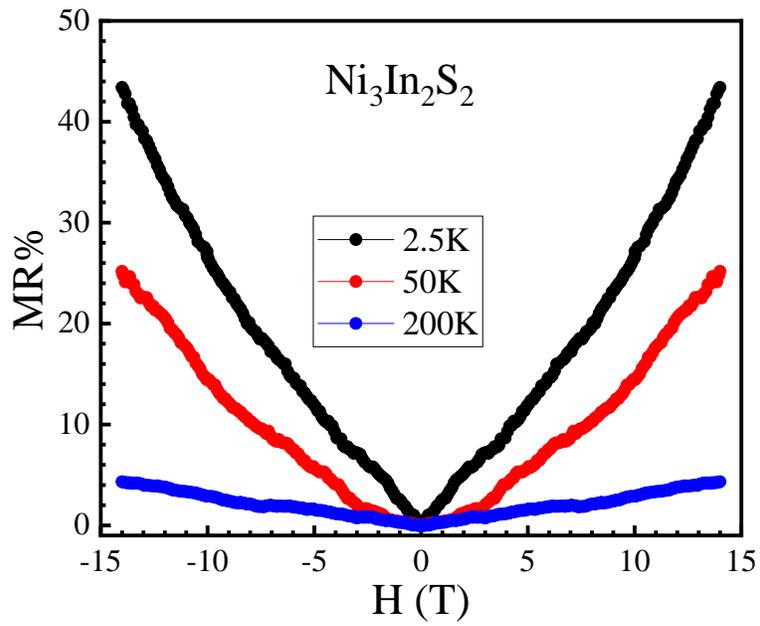

Fig.6:

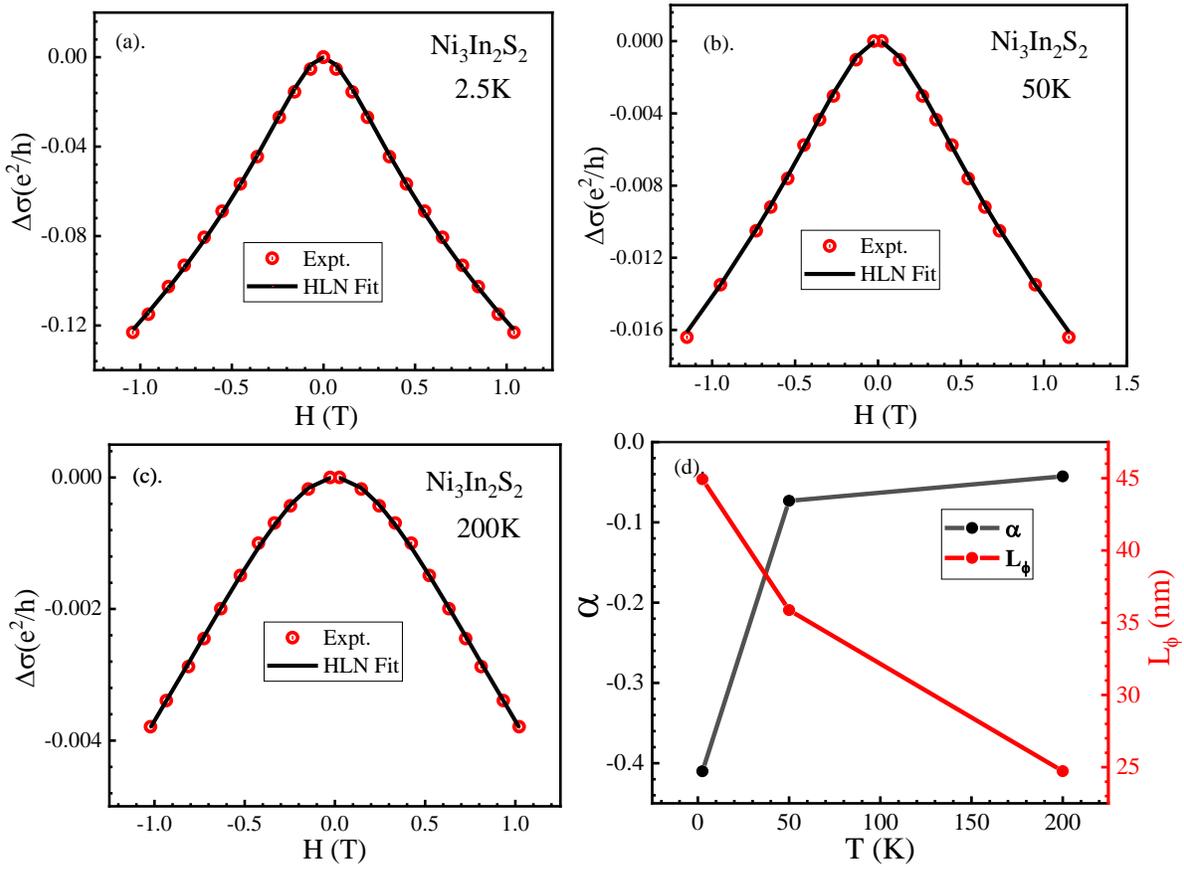

Fig. 7:

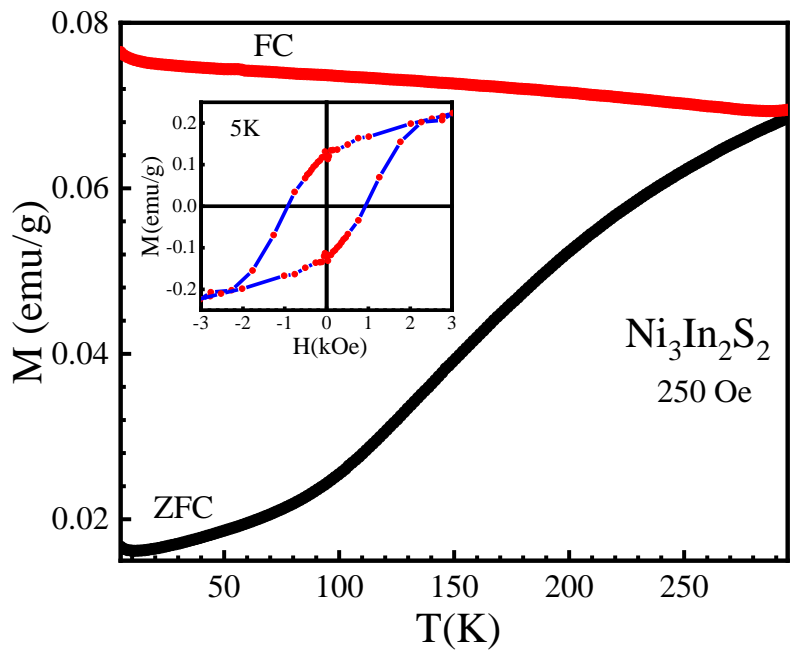